\documentstyle[12pt]{article}
\begin{document}

\author{Minu Joy$^{\dagger }$ and V. C. Kuriakose$^{\ddagger }$ \\
Department of Physics, Cochin University of Science and Technology, \\
India-682 022}
\title{First order phase transitions in a \\
Bianchi type-I universe }
\date{}
\maketitle

\begin{abstract}
Considering the theory of induced gravity coupled to matter fields, taking
the $\phi ^6$ interaction potential model we evaluate the one-loop effective
potential in a (3+1)dimensional Bianchi type-I spacetime. It is proved that
the $\phi ^6$ theory can be regularised in (3+1)dimensional curved
spacetime. We evaluate the finite temperature effective potential and study
the temperature dependence of phase transitions. The nature of phase
transitions in the early universe is clarified to be of first order. The
effects of spacetime curvature and arbitrary field coupling on the phase
transitions in the early universe are also discussed.

$\dagger $e-mail: minujoy@cusat.ac.in

$^{\ddagger }$e-mail: vck@cusat.ac.in

PACS number(s): 04.62.+v, 11.10.Gh, 11.10.Wx, 11.15.Ex

\newpage\ 
\end{abstract}

\begin{center}
{\bf I. INTRODUCTION}
\end{center}

In the early universe, symmetries that are spontaneously broken today were
restored and during the evolution of the universe there were phase
transitions, perhaps many, associated with the spontaneous breakdown of
symmetries(SSB) [1]. During such a phase transition it is possible for the
field to acquire nonzero vacuum expectation values. In general, a symmetry
breaking phase transition can be first or second order. For a first order
phase transition the change in $\phi \ $in going from one phase to the other
must be discontinuous, while for a second order transition there is no
barrier at the transition point and the transition occurs smoothly. Of
particular interest for cosmology is the nature of phase transition, whether
it is first order or not. If the phase transition is strongly first order,
the Universe may be dominated by the vacuum energy and undergo a period of
inflation. In this case, the transition proceeds by the nucleation of
bubbles of the true vacuum. If the phase transition\ is higher order, or
weakely first order, thermal fluctuations may drive the transition.

Quantum fields have profound influence on the dynamical behaviour of the
early universe [2-4]. Quantum field theory in an external classical
gravitational field is usually regarded as a first step towards a more
complete theory of quantum gravity [5]. At high energies the quantum matter
fields are free from all the interactions except the conformal one with an
external metric. The requirement of the conformal invariance is especially
important for the scalar field, as it fixes the value of the non-minimal
parameter of the scalar curvature interaction to the conformal value. The
effect of quantum conformal factor leads to a first order phase transition
induced by curvature where the scalar field plays the role of order
parameter [6-8].

The influence of quantum fields and the gravitational effects on the
cosmological phase transitions have been investigated by many authors
[9-12]. From analysis based on the one loop renormalised effective potential
it is concluded that the scalar gravitational coupling $\xi \ $and the
magnitude of the scalar curvature R crucially determine the fate of
symmetry. At the classical level the scalar curvature acts as an effective
mass of the field and thus influence the phase transition of the system. The
effect of anisotropy on the static spacetimes like Mixmaster or Taub
Universe on the process of symmetry breaking and restoration has also been
discussed [13,14].

In the present work we discuss the quantum field effects on phase transition
and the temperature dependence of phase transition for a $\phi ^6$ theory in
a Bianchi type-I universe. This is the most general model for a
self-interacting scalar field exhibiting $\phi \longrightarrow -\phi $
symmetry.$\ $Self interactions upto $\phi ^6$ exhibit three lowest levels
well separated from the rest [15]. Boyanovsky and Masperi have shown that
the nature of phase transitions associated with such a field system may be
of first order or second order depending on the relative depths of the wells
and the strength of coupling.

One of the simplest models of an anisotropic universe that describes a
homogeneous and a spatially flat universe is the Bianchi type-I cosmology.
Unlike the FRW model which has the same scale factor for each of the three
spatial directions, the Bianchi type-I cosmology has a differrent scale
factor in each direction, which produces an anisotropy in expansion.
Futamase has considered the effective potential in a Bianchi type-I universe
[16] which reduces to the spatially flat Robertson-Walker model for zero
anisotropy. Huang has discussed the fate of symmetry in a Bianchi type-I
universe using an adiabatic approximation for a massless field with
arbitrary coupling to gravity [17]. Berkin has also calculated the effective
potential in a Bianchi type-I universe, for a scalar field having arbitrary
mass and coupling to gravity [18].

$\phi ^6$ model is known to be nonrenormalisable in (3+1)dimensional flat
spacetime. Nonrenormalizability of the field theory does not mean that the
theory is not interesting and it does not mean, ofcourse, that finite
renormalised prescription for the calculation of one-loop effective
potential does not exist[19]. Using the present $\phi ^6$ model we have
obtained a finite expression for the one-loop effective potential. The
present calculations show that the $\phi ^6$ model can be regularised using
the effective potential method in (3+1)dimensional curved spacetime. This
paper is organised in the following way. In section II we evaluate the
one-loop effective potential for $\phi ^6$ theory in a $\left( 3+1\right) $
dimensional Bianchi type-I spacetime with small anisotropy and discuss the
properties of quantum field corrections to the symmetry breaking or
restoration. The finite temperature effects on the phase transitions of
early universe are discussed in section III and the nature of phase
transitions is examined in section IV. The crucial dependence of phase
transitions of the early universe on spacetime curvature and the
gravitational-scalar coupling is made clear in section V. Section VI is
devoted to discussions and conclusions of the present calculations.

{\bf II. QUANTUM FIELD EFFECTS ON SYMMETRY\ BREAKING AND RESTORATION\ IN
BIANCHI\ TYPE-I\ SPACETIME}

The Lagrangian density $\pounds ${\sl \ }describing a massive{\sl \ } self
interacting{\sl \ } scalar field{\sl \ $\phi $ }coupled{\sl \ }arbitrarily
to the gravitational back ground is,

\begin{equation}
\label{1}\pounds =\sqrt{-g}\left\{ \frac 12[g^{\mu \nu }\partial _\mu \phi
\partial _\nu \phi -\xi R\phi ^2]-\frac 12\lambda ^2\phi ^2(\phi
^2-m/\lambda )^2\right\} 
\end{equation}
{\sl \ }where, the classical potential corresponding to this Lagrangian is,

\begin{equation}
\label{2}V(\phi )=\frac 12\xi R\phi ^2+\frac 12\lambda ^2\phi ^2(\phi
^2-m/\lambda )^2 
\end{equation}

This Lagrangian exhibits $\phi \longrightarrow -\phi $ symmetry. The
equation of motion associated with the Lagrangian(1) is,\ 

\begin{equation}
\label{3}g^{\mu \nu }\nabla _\mu \nabla _\nu \phi +(m^2+\xi R)\phi -4\kappa
\phi ^3+3\lambda ^2\phi ^5=0 
\end{equation}
{\sl \ }in which we put $m$$\lambda =\kappa .$ We can write, 
\begin{equation}
\label{4}\phi =\phi _c+\phi _q 
\end{equation}
{\sl \ }where $\phi _c$\ is the classical field and $\phi _q$\ is\ a quantum
field with vanishing vacuum expectation value, $<\phi _q>$ $=0.\ $
Introducing renormalised parameters,

\begin{equation}
\label{5}
\begin{array}{c}
m^2=m_r^2+\delta m^2,\ \xi =\xi _r+\delta \xi , \\  
\\ 
\kappa =\kappa _r+\delta \kappa ,\qquad \ \lambda =\lambda _r+\delta \lambda 
\end{array}
\end{equation}
the field equaton for the classical field $\phi _c$ becomes,{\sl \ } {\sl \ }

\begin{equation}
\label{6}
\begin{array}{c}
g^{\mu \nu }\nabla _\mu \nabla _\nu \phi _c+[(m_r^2+\delta m^2)+(\xi
_r+\delta \xi )R]\phi _c \\  
\\ 
-4(\kappa _r+\delta \kappa )\phi _c^3-12(\kappa _r+\delta \kappa )\phi
_c<\phi _q^2> \\  
\\ 
+\ 3(\lambda _r^2+\delta \lambda ^2)\phi _c^5+30(\lambda _r^2+\delta \lambda
^2)\phi _c^3<\phi _q^2> \\  
\\ 
+15(\lambda _r^2+\delta \lambda ^2)\phi _c<\phi _q^4>\ =0 
\end{array}
\end{equation}
and to the one loop quantum effect, the field equation for the quantum field 
$\phi _q$ is,

\begin{equation}
\label{7}g^{\mu \nu }\nabla _\mu \nabla _\nu \phi _q+(m_r^2+\xi R)\phi
_q-12\kappa _r\phi _c^2\phi _q+15\lambda _r^2\phi _c^4\phi _q=0 
\end{equation}
The effective potential V$_{eff}\ $is given by,

\begin{equation}
\label{8}
\begin{array}{c}
V_{eff}=\frac 12[(m_r^2+\delta m^2)+(\xi _r+\delta \xi )R][\phi _c^2+<\phi
_q^2>] \\  
\\ 
-(\kappa _r+\delta \kappa )\phi _c^4-6(\kappa _r+\delta \kappa )\phi
_c^2<\phi _q^2> \\  
\\ 
-(\kappa _r+\delta \kappa )<\phi _q^4>+\frac 12(\lambda _r^2+\delta \lambda
^2)\phi _c^6 \\  
\\ 
+ 
\frac{15}2(\lambda _r^2+\delta \lambda ^2)\phi _c^4<\phi _q^2> \\  \\ 
+ 
\frac{15}2(\lambda _r^2+\delta \lambda ^2)\phi _c^2<\phi _q^4> \\  \\ 
+\frac 12(\lambda _r^2+\delta \lambda ^2)<\phi _q^6> 
\end{array}
\end{equation}
To make V$_{eff}$ finite, the following renormalisation conditions are used,

\begin{equation}
\label{9}
\begin{array}{c}
m_r^2=\left( 
\frac{\partial ^2V_{eff}}{\partial \phi _c^2}\right) _{\phi _c=R=0} \\  \\ 
\xi _r=\left( 
\frac{\partial ^3V_{eff}}{\partial R\partial \phi _c^2}\right) _{\phi
_c=R=0} \\  \\ 
\kappa _r=\left( 
\frac{\partial ^4V_{eff}}{\partial \phi _c^4}\right) _{\phi _c{\ }=R=0} \\  
\\ 
{\sl \lambda _r^2=\left( \frac{\partial ^6V_{eff}}{\partial \phi _c^6}
\right) }_{\phi _c=R=0} 
\end{array}
\end{equation}
\ To evaluate $<\phi _q^2>,$ $<\phi _q^4>,$ and $<\phi _q^6>$ we adopt the
canonical quantisation relations: 
\begin{equation}
\label{10}[\phi _q(t,x),\phi _q(t,y)]=[\pi _q(t,x),\pi _q(t,y)]=0; 
\hspace{1.3cm}[\phi _q(t,x),\pi _q(t,y)]=i\delta ^3(x-y) 
\end{equation}
where the conjugate momentum $\pi _q$ is defined by $\pi _q=\frac{\partial
\pounds }{\partial (\partial _i\phi )}$ . Due to the space homogeneity we
expand the quantum field $\phi _q$ by the summation or integration over
modes in the form,

\begin{equation}
\label{11}\phi _q(t,x)=C^{-1/2}(t)\int d\mu (k)[a_k\chi
_k(t)y_k(x)+a_k^{+}\chi _k^{*}(t)y_k^{*}(x)] 
\end{equation}
where $y_k(x)$ is a normalised eigen function of the spatial part of field
equation, while $\chi _k(t)$ is that of the time part. An explicit
functional form of the mode solutions $\chi _k(t)$ and $y_k(x)$ can only be
found after specifying the background spacetime.

We consider a (3+1) dimensional Bianchi type-I spacetime with small
anisotropy which has the line element

\begin{equation}
\label{12}
\begin{array}{c}
ds^2=C(\eta )d\eta ^2-a_1^2(\eta )dx^2-a_2^2(\eta )dy^2-a_3^2(\eta )dz^2 \\  
\\ 
C=(a_1a_2a_3)^{2/3} 
\end{array}
\end{equation}
In this model the mode function can be written in the separated form as\\ u$
_k=C^{-1/2}(2\pi )^{-3/2}\exp (i\kappa .x)\chi _k(\eta )$ and then

\begin{equation}
\label{13}
\begin{array}{c}
<\phi _q^2(\eta )>=\frac 1{8\pi ^3C(\eta )}\int d^3k\chi _k(\eta )\chi
_k^{*}(\eta ), \\  
\\ 
<\phi _q^4(\eta )>=\frac 1{64\pi ^6C^2(\eta )}\int d^3k(\chi _k(\eta )\chi
_k^{*}(\eta ))^2and \\  
\\ 
<\phi _q^6(\eta )>=\frac 1{512\pi ^9C^3(\eta )}\int d^3k(\chi _k(\eta )\chi
_k^{*}(\eta ))^3 
\end{array}
\end{equation}
The wave equation Eq. (7) will then lead to

\begin{equation}
\label{14}\stackrel{\cdot \cdot }{\chi }+\left\{ C\left[ m_r^2+(\xi _r-\frac
16)R-12\kappa _r\phi _c^2+15\lambda _r^2\phi _c^4+\sum\limits_i\frac{k_i^2}{
a_i^2}\right] +Q\right\} \chi _k=0 
\end{equation}

where the spacetime curvature function R and the anisotropic function Q are

\begin{equation}
\label{15}
\begin{array}{c}
R=6C^{-1}( 
\stackrel{\bullet }{H}+H^2+Q)\hspace{3cm}H=\sum\limits_ih_i \\  \\ 
h_i=\frac{\stackrel{\cdot }{a_i}}{a_i}\hspace{3cm}Q=\frac
1{36}\sum\limits_{i<j}(h_i-h_j)^2 
\end{array}
\end{equation}

When the metric is slowly varying Eq. (14) possesses WKB approximation
solution:

\begin{equation}
\label{16}
\begin{array}{c}
\chi _k=(2W_k)^{-\frac 12}\exp (-i\int d\eta W_k) \\ 
where \\ 
W_k=\left\{ C\left[ m_r^2+(\xi _r-\frac 16)R-12\kappa _r\phi _c^2+15\lambda
_r^2\phi _c^4+\sum\limits_i\frac{k_i^2}{a_i^2}\right] +Q\right\} ^{\frac 12} 
\end{array}
\end{equation}
Substituting the above solution in Eq. (13):

\begin{equation}
\label{17}
\begin{array}{c}
<\phi _q^2>=\frac 1{16\pi ^3C(\eta )}\int d^3k\left\{ C\left[ m_r^2+(\xi
_r-\frac 16)R-12\kappa _r\phi _c^2+15\lambda _r^2\phi _c^4+\sum\limits_i 
\frac{k_i^2}{a_i^2}\right] +Q\right\} ^{-\frac 12} \\  \\ 
=\frac 1{16\pi }\{\Lambda ^2+\frac 12\left[ m_r^2+(\xi _r-\frac
16)R-12\kappa _r\phi _c^2+15\lambda _r^2\phi _c^4+\frac QC\right] \times \\  
\\ 
\hspace{6cm}\left[ 1+\ln \left[ \frac{m_r^2+(\xi _r-\frac 16)R-12\kappa
_r\phi _c^2+15\lambda _r^2\phi _c^4+\frac QC}{4\Lambda ^2}\right] \right] \} 
\end{array}
\end{equation}

and similarly,

\begin{equation}
\label{18}
\begin{array}{l}
<\phi _q^4>=\frac \Lambda {128\pi ^4C^{\frac 32}}\{1- 
\frac{\left[ m_r^2+(\xi _r-\frac 16)R-12\kappa _r\phi _c^2+15\lambda
_r^2\phi _c^4+\frac QC\right] ^{\frac 12}}\Lambda \times \\  \\ 
\hspace{6cm}\arctan \frac \Lambda {\left[ m_r^2+(\xi _r-\frac 16)R-12\kappa
_r\phi _c^2+15\lambda _r^2\phi _c^4+\frac QC\right] ^{\frac 12}}\} 
\end{array}
\end{equation}

where we have introduced a momentum cut-off $\Lambda $ to regularise the
k-integration. From the renormalisation conditions given by Eq. (9),\ the
renormalisation counter terms are evaluated as,

\begin{equation}
\label{19}
\begin{array}{c}
\delta m^2= 
\frac{3(\kappa _r+\delta \kappa )}{4\pi }\left[ \Lambda ^2+\frac 12\left(
m_r^2+\frac QC\right) \left( 1+\ln \left[ \frac{m_r^2+\frac QC}{4\Lambda ^2}
\right] \right) \right] \\  \\ 
\hspace{1.5cm}-\frac{15(\lambda _r^2+\delta \lambda ^2)}{128\pi ^4C^{\frac
32}}\left[ \Lambda -\left( m_r^2+\frac QC\right) ^{\frac 12}\arctan \left[
\frac \Lambda {(m_r^2+\frac QC)^{\frac 12}}\right] \right] 
\end{array}
\end{equation}

\begin{equation}
\label{20}
\begin{array}{l}
\delta \xi = 
\frac{3(\kappa _r+\delta \kappa )}{8\pi }\left( \xi _r-\frac 16\right)
\left[ 2+\ln \left[ \frac{m_r^2+\frac QC}{4\Lambda ^2}\right] \right] \\  \\ 
\hspace{1.5cm}+\frac{15(\lambda _r^2+\delta \lambda ^2)}{256\pi ^4C^{\frac
32}}\left( \xi _r-\frac 16\right) \left[ \frac 1{(m_r^2+\frac QC)^{\frac
12}}\arctan \left[ \frac \Lambda {(m_r^2+\frac QC)^{\frac 12}}\right] -\frac
\Lambda {(m_r^2+\frac QC+\Lambda ^2)}\right] 
\end{array}
\end{equation}

\begin{equation}
\label{21}\delta \kappa =-\kappa _r-\frac{\lambda _r^2}{60\left[ \frac{
45\lambda _r^2}{4\pi }\left( 2+\ln \left[ \frac{m_r^2+\frac QC}{4\Lambda ^2}
\right] \right) +\frac{54\kappa _r^2}{\pi (m_r^2+\frac QC)}\right] } 
\end{equation}

\begin{equation}
\label{22}
\begin{array}{c}
\begin{array}{l}
\delta \lambda _r^2=-\lambda _r^2 \\  
\\ 
\hspace{1.5cm}+\frac{8\left[ -\lambda _r^2\pi +27\kappa _r\lambda _r^2\left(
2+\ln \left[ \frac{m_r^2+\frac QC}{4\Lambda ^2}\right] \right) +\frac{
135\kappa _r^3}{\left( m_r^2+\frac QC\right) }\right] }{225\left[ \Lambda
^2+\frac 12(m_r^2+\frac QC)\left( 1+\ln \frac{m_r^2+\frac QC}{4\Lambda ^2}
\right) +\frac{3\kappa _r}{16\pi ^3}\left( \frac 1{(m_r^2+\frac QC)^{\frac
12}}\arctan \left[ \frac \Lambda {(m_r^2+\frac QC)^{\frac 12}}\right] -\frac
\Lambda {m_r^2+\frac QC+\Lambda ^2}\right) \right] } 
\end{array}
\\  
\\  
\\ 
\hspace{1.5cm}\frac 1{\left[ \frac{45\lambda _r^2}{4\pi }\left( 2+\ln \left[ 
\frac{m_r^2+\frac QC}{4\Lambda ^2}\right] \right) +\frac{54\kappa _r^2}{\pi
(m_r^2+\frac QC)}\right] } 
\end{array}
\end{equation}
Substituting the renormalisation counterterms, we find $\frac{\partial
V_{eff}}{\partial \phi _c}$ obtained from Eq. (8) as,

\begin{equation}
\label{23}
\begin{array}{l}
\frac{\partial V_{eff}}{\partial \phi _c}=(m_r^2+\xi _rR)\phi _c \\  \\ 
\hspace{1.5cm}+\frac{\kappa _r(\frac{n\pi }2)}{100\pi ^3C\frac 32\left[
(m_r^2+\frac QC)+\frac{3\kappa _r}{16\pi ^3(m_r^2+\frac QC)^{\frac 12}}( 
\frac{n\pi }2)\right] } \\  \\ 
\hspace{3cm}\left\{ (m_r^2+\frac QC)^{\frac 12}-\left[ m_r^2+(\xi _r-\frac
16)R-12\kappa _r\phi _c^2+15\lambda _r^2\phi _c^4+\frac QC\right] \right\}
\phi _c \\  \\ 
\hspace{1.5cm}+\left[ -\frac{(\xi _r-\frac 16)}{900}+\frac{(\xi _r-\frac
16)\kappa _r(\frac{n\pi }2)}{200\pi ^3(m_r^2+\frac QC)^{\frac 12}\left[
(m_r^2+\frac QC)+\frac{3\kappa _r}{16\pi ^3(m_r^2+\frac QC)^{\frac 12}}( 
\frac{n\pi }2)\right] }\right] R\phi _c \\  \\ 
\hspace{1.5cm}+\frac{2\kappa _r\left[ m_r^2+(\xi _r-\frac 16)R-12\kappa
_r\phi _c^2+15\lambda _r^2\phi _c^4+\frac QC\right] }{25\left[ (m_r^2+\frac
QC)+\frac{3\kappa _r}{16\pi ^3(m_r^2+\frac QC)^{\frac 12}}(\frac{n\pi }
2)\right] } \\  \\ 
\hspace{4cm}\left[ 1+\ln \left( m_r^2+(\xi _r-\frac 16)R-12\kappa _r\phi
_c^2+15\lambda _r^2\phi _c^4+\frac QC\right) \right] \phi _c^3 \\  \\ 
\hspace{1.5cm}+\frac{32\kappa _r\pi }{375\left[ (m_r^2+\frac QC)+\frac{
3\kappa _r}{16\pi ^3(m_r^2+\frac QC)^{\frac 12}}\left( \frac{n\pi }2\right)
\right] }\phi _c^5\hspace{3cm}where\hspace{.3cm}n=1,2,3.... 
\end{array}
\end{equation}

The above equation shows that we can obtain finite expression for the one
loop effective potential using this $\phi ^6$ model in (3+1)dimensional
Bianchi Type I spacetime. Thus it is clear that the $\phi ^6$ theory in
(3+1)dimension can be regularised in a curved anisotropic spacetime using
the effective potential method. It is to be noted that once we let the
anisotropy in the above equation be zero, our result is consistent with that
of the symmetric homogeneous case.

Now we are in a position to investigate the gravitational and quantum field
effects on the cosmological phase transitions. This can be done by
considering the case $\phi _c\longrightarrow 0$. In the case of conformal
coupling ($\xi _r=\frac 16)$ or vanishing scalar curvature(R =0) we have,

\begin{equation}
\label{24}\left( \frac{\partial V_{eff}}{\partial \phi _c}\right) _{\phi
_c\longrightarrow 0}\sim m_r^2\phi _c
\end{equation}
which shows that in such situations, the one-loop quantum correction does
not change the fate of symmetry. For the other cases, we can find from the
above equations that only for some suitable values of scalar gravitational
coupling could the symmetry be radiatively broken or restored.

The perturbative method of calculating the effective potential can be
improved by using Renormalisation Group(RG) approach[20]. Such RG improved
effective potential can be calculated in curved spacetime too [21]. The
condition expressing the independence of the effective potential from the
renormalisation point leads to Renormalisation Group Equation(RGE) [6]. This
property in renormalisable theories may be used for construction of famous
RG improved effective potential, which is much more exact than one
loop-effective potential, because it takes into account of all orders of the
perturbation theory. However, unlike to such multiplicatively renormalisable
theories RG improved potential will not give leading log approximation in
the present $\phi ^6$ model as the theory is not multiplicatively
renormalisable.

\begin{center}
{\bf III. FINITE TEMPERATURE BEHAVIOUR}
\end{center}

The evolution of particles in vacuum and in a thermal bath are very
different. Similarly, the nature of evolution of field changes when coupled
to a thermal bath.Under certain conditions, the changes may be absorbed in a
temperature dependent potential, the finite temperature effective potential.
The temperature dependence of finite temperature effective potential in
quantum field theory leads to phase transitions in the early universe [22].
In this case the vaccum expectation value is replaced by the thermal average 
$<\phi >_T=\sigma _T$ taken with respect to a Gibbs ensemble [1].

In this section we evaluate the effective potential at finite temperature
and show that the symmetry breaking present in the model can be removed if
the temperature is raised above a certain value called the critical
temperature.

Considering the same Lagrangian density as above, the zero loop effective
potential is temperature independent, 
\begin{equation}
\label{25}V_0(\sigma )=\frac 12\xi R\sigma ^2+\frac 12\lambda ^2\sigma
^2(\sigma ^2-m/\lambda )^2
\end{equation}
The one loop approximation to finite temperature effective potential has
been computed by many authors [23-26] and is given by,

\begin{equation}
\label{26}
\begin{array}{c}
V_1^\beta (\sigma )=\frac 1{2\beta }\sum_n\int 
\frac{d^3k}{(2\pi )^3}\ln (k^2-M^2) \\  \\ 
\hspace{1.92cm}=\frac 1{2\beta }\sum_n\int \frac{d^3k}{(2\pi )^3}\ln (\frac{
-4\pi ^2n^2}{\beta ^2}-E_M^2) 
\end{array}
\end{equation}

\begin{equation}
\label{27}
\begin{array}{c}
where,E_M^2=k^2+M^2, \\  
\\ 
M^2=m^2+\xi R-12\lambda m\sigma ^2+15\lambda ^2\sigma ^4
\end{array}
\end{equation}
The sum over n diverges; it may be evaluated by the method of Dolan and
Jackiw [23] and we get,

\begin{equation}
\label{28}
\begin{array}{l}
V_1^\beta (\sigma )=\int 
\frac{d^3k}{(2\pi )^3}\left[ \frac{E_M}2+\frac 1\beta \ln (1-e^{-\beta
E})\right] \\  \\ 
\hspace{1.22cm}=V_1^0(\sigma )+\bar V_1^\beta (\sigma ) 
\end{array}
\end{equation}
where, 
\begin{equation}
\label{29}V_1^0(\sigma )=\int \frac{d^3k}{(2\pi )^3}\frac{E_M}2, 
\end{equation}
and

\begin{equation}
\label{30}
\begin{array}{c}
\bar V_1^\beta (\sigma )=\int 
\frac{d^3k}{(2\pi )^3}\frac 1\beta \ln (1-e^{-\beta E}) \\  \\ 
\hspace{5cm}=\frac 1{4\pi \beta ^3}\int\limits_0^\infty xdx\ln \left[ 1-\exp
-(x^2+\beta ^2M^2)^{1/2}\right] 
\end{array}
\end{equation}
where we put x$^2/\beta ^2=E_M^2-M^2.$ The integral may be evaluated by
expanding $\bar V_1^\beta (\sigma )$ as a Taylor series and in the high
temperature limit we find that 
\begin{equation}
\label{31}V_1^\beta (\sigma )=\frac{-\pi ^2}{90\beta ^4}+\frac{M^2}{24\beta
^2}-\frac{M^3}{12\pi \beta }-\frac{M^4}{64\pi ^2}\ln M^2\beta ^2 
\end{equation}

The critical temperature in the present case is

\begin{equation}
\label{32}T_c=\left[ \frac{\left( m^2+\xi R\right) }{\lambda m}\right]
^{\frac 12} 
\end{equation}

The symmetry breaking present in the model can be removed if the temperature
is raised above a certain value called the critical temperature. The order
parameter of the theory is temperature dependent.

\begin{center}
{\bf IV. NATURE OF PHASE TRANSITION}
\end{center}

The characteristic of a first order phase transition is the existence of a
barrier between the symmetric and the broken phase [27]. The temperature
dependence of V$_{eff}$ for a first order phase transition obtained using
the present $\phi ^6$ model is shown in Fig.1(a- e). When T$\gg$T$_c,$ the effective potential attains a minimum at $\sigma =0,
$ which corresponds to the completely symmetric case. When the temperature
decreases, a global minimum appears at $\sigma =0$ and two local minima at $
\sigma \neq 0,$ which shows the existence of a barrier between the global
and local minima$.$ At T=T$_c$, all the minima are degenerate, that means
the symmetry is broken. For T<T$_c$ the minima at $\sigma \neq 0
$ becomes the global one. If for T$\leq T_c$ the extremum at $\sigma =0$
remains a local minimum, there must be a barrier between the minimum at $
\sigma =0$ and at $\sigma \neq 0.$ Therefore the change in $\sigma $ in
going from one phase to the other must be discontinuous, indicating a first
order phase transition. The phase transition starts at T$_c$ by tunnelling,
however, if the barrier is high enough the tunnelling effect is very small
and the phase transition does effectively starts at a temperature T$\ll $T$_c
$ [28]. This shows that the present model can describe first order phase
transitions which might have taken place during the evolution of the early
universe.\ 

\begin{center}
{\bf V. DEPENDENCE ON SCALAR CURVATURE R AND SCALAR-GRAVITATIONAL COUPLING }$
{\bf \xi }$
\end{center}

Using this $\phi ^6$ model, it is proved that the curvature can restore
broken symmetries for a wide range of parameters from conformal to near
minimal couplings, even if the temperature is below critical temperature.
Fig. 2 clearly shows that the first order phase transition takes place as R
changes.

The saclar-gravitational coupling constant $\xi $ is found to play a crucial
role in symmetry breaking phase transitions. Classicaly, a positive $\xi $
restores symmetry, while the opposite effects are found for negative
coupling [18]. Quantum effects depend on the value of $\xi $ relative to the
conformal value $\frac 16.$ The present calculations show that the symmetry
is restored as the scalar coupling constant $\xi $ is increased. This phase
transition, induced by the coupling constant $\xi $ is also found to be of
first order. It is clear from Fig. 3 that there is a barrier between the
symmetric and broken phases.

\begin{center}
{\bf VI. DISCUSSIONS AND CONCLUSIONS}
\end{center}

According to renormalizability considerations, degree of the interaction
potential can not be higher than four in (3+1)dimension [19]. The present
calculations show that the $\phi ^6$ theory in (3+1)dimension can be
regularised in curved spacetime and one can obtain finite expression for the
one loop effective potential. In this paper we closely examine and verify
the temperature dependence of phase transitions in the early universe and
verify their nature to be of first order as the transition is found to be
discontinuous.

In most of the works on cosmological phase transitions, the coupling to the
background gravitational field is ignored. One deals with the Quantum field
theory in flat spacetime at finite temperature and the expansion of the
Universe serves only to decrease the temperature. However, at sufficiently
early times the spacetime curvature can be expected to be important. Many
authors have argued that such effects may be important in the context of
cosmological phase transitions in Grand Unified models [19,29-32]. Vilenkin
and Ford have shown that spacetime curvature can drastically change the
behaviour of the system [33]. O'Connor and co-workers have confirmed the
effect of spacetime curvature and arbitrary field coupling on the phase
transitions of the early universe [34]. Janson [35], Grib and Mosteparenko
[36] and Madsen [37] have independently shown that the interaction with the
external gravitational field may lead to SSB. The present work proves that
the phase transition taking place during such a SSB is first order. It is
found that for $\xi =0$ or $R=0$ the system remains in the symmetry broken
state for all values of T$\leq T_c$. As the temperature is increased above T$
_c,$ the symmetry is restored depending on the values of $\xi $ and $R$ also$
.$ It is also found that symmetry can be restored either by increasing the
value of $\xi $ or by increasing the value of $R$ keeping the temperature
constant. This shows that the scalar-gravitational coupling and the scalar
curvature do play a crucial role in determining the nature of phase
transitions took place in the early universe.

These results may be useful for the study of quantum thermal processes in
the early universe. To examine the symmetry behaviour of the early universe
closely one should take into consideration the effects of spacetime
curvature and finite temperature corrections in their full rights.

\begin{center}
{\bf ACKNOWLEDGEMENTS}
\end{center}

The authors are thankful to the referee for the valuable comments. We thank
Prof. T. Padmanabhan and Prof. Maria Lombardo for valuable discussions. One
of us (MJ) would like to thank UGC, N. Delhi for financial support in the
form of a JRF. VCK\ acknowledges Associateship of IUCAA, Pune.

\newpage\ {\bf Figure Captions}

\begin{enumerate}
\item  Fig. 1a : The behaviour of finite temperature effective potential as
a function of $\sigma $ for fixed $m=0.9371$, $\lambda =0.008,$ $R=11.2,\xi
=1.6$ and $T=50$ such that ${\bf T>>T}_c,$for which the symmetry is
completely restored.

\item  Fig. 1b :The behaviour of finite temperature effective potential as a
function of $\sigma $ for fixed $m=0.9371$, $\lambda =0.008,$ $R=0.8,\xi
=0.145$ and $T=10.15$ such that ${\bf T>T}_c{\bf .}$

\item  Fig. 1c : The behaviour of finite temperature effective potential as
a function of $\sigma $ for fixed $m=0.9371$, $\lambda =0.008,$ $R=0.35,\xi
=0.004$ and $T=8.69$ such that ${\bf T=T}_c{\bf .}$

\item  Fig. 1d : The behaviour of finite temperature effective potential as
a function of $\sigma $ for fixed $m=0.9371$, $\lambda =0.008,$ $R=0.31,\xi
=-0.22$ and $T=5$ such that ${\bf T<T}_c{\bf .}$

\item  Fig. 1e : The behaviour of finite temperature effective potential as
a function of $\sigma $ for fixed $m=0.9371$, $\lambda =0.008,$ $R=0.3,\xi
=-0.3$ and ${\bf T=0}.$

\item  Fig. 2 : The behaviour of finite temperature effective potential as a
function of $\sigma $ for fixed $m=0.9371$, $\lambda =0.008,$ $\xi =0.1$ and
T$=1.$ Starting from top the curves corresponds to the following values of
the curvature : R=15,4,2.5,0.5,0.001,-0.9.

\item  Fig. 3 : The behaviour of finite temperature effective potential as a
function of $\sigma $ for fixed $m=0.9371$, $\lambda =0.008,$ $R=0.3$ and $
T=3.$ Starting from top the curves corresponds to the following values of
the curvature:$\xi $ = 6.5,2.3,1.25,0.01,-0.3,-0.7.
\end{enumerate}


\begin{thebibliography}{99}
\bibitem{1}  A. D. Linde, Rep. Prog. Phys. {\bf 42,} 389 (1979).

\bibitem{2}  N. D. Birrel and P. C. W. Davies,{\it \ Quantum Fields in
Curved Space (}Cambridge University Press, Cambridge, England, 1982).

\bibitem{3}  D. G. C. McKeon and G. Tsoupros, Class. Quantum Grav. {\bf 11, }
73 (1994).

\bibitem{4}  S. A. Ramsey and B. L. Hu, Phys. Rev. D {\bf 56, }661 (1997).

\bibitem{5}  E. Elizalde, S. Leseduarte, S. D. Odintsov and Yu I. Shil'nov,
Phys. Rev. D {\bf 53, }1917 (1996).

\bibitem{6}  I. L. Buchbinder and S. O. Odinstov, Class. Quantum Grav. {\bf %
2, }721 (1985).

\bibitem{7}  G. Cognola and I. L. Shapiro, Class. Quantum Grav. {\bf 15, }
787 (1998).

\bibitem{8}  T. Inagaki, T. Muta and S. D. Odintsov, Mod. Phys. Lett {\bf A
8, }2117 (1993).

\bibitem{9}  W. H. Huang, Class. Quantum Grav. {\bf 10, }2021 (1993).

\bibitem{10}  H. Ford and D. J. Toms, Phys. Rev. D {\bf 25, }1510 (1982). 
{\bf \ } {\bf \ }

\bibitem{11}  B. L. Hu and Y. Zhang, Phys. Rev. D {\bf 37, }2125 (1988).

\bibitem{12}  A. Ringwald, Phys. Rev. D {\bf 36, }2598 (1987).

\bibitem{13}  R. Critchley and J. S. Dowker, J. Phys. A: Math. Gen. {\bf 15}
, 157 (1982).

\bibitem{14}  T. C. Shen, B. L. Hu and D. J. O'Connor, Phys. Rev. D {\bf 31, 
}2401 (1985).

\bibitem{15}  D. Boyanovsky and L. Masperi, Phys. Rev. D {\bf 21, }1550
(1980).

\bibitem{16}  T. Futamase, Phys. Rev. D {\bf 29}, 2783 (1984).

\bibitem{17}  W. H. Huang, Phys. Rev. D {\bf 42}, 1282 (1990).

\bibitem{18}  A. L. Berkin, Phys. Rev. D {\bf 46}, 1551 (1992).

\bibitem{19}  I. L. Buchbinder, S. D. Odinstov and I. L. Shapiro, {\it 
Effective Action in Quantum Gravity (}IOP Publishing {\it \ }Ltd, 1992).

\bibitem{20}  S. Coleman and E. Weinberg, Phys. Rev. D {\bf 7}, 1888 (1973).

\bibitem{21}  E. Elizalde and S. D. Odinstov, Z. Phys. C {\bf 64}, 699
(1994). 

\bibitem{22}  R. Brandenberger, {\it Proceedings of the 1991 Summer School
on High Energy Physics and Cosmology,ICTP,Trieste(World
Scientific,Singapore).}

\bibitem{23}  L. Dolan and R. Jackiw, Phys. Rev. D {\bf 9}, 2904 (1974).

\bibitem{24}  K. Babu Joseph and V. C. Kuriakose, J. Phys. A: Math. Gen. 
{\bf 15, }2231 (1982).

\bibitem{25}  Moss, Toms and Wright, Phys. Rev. D {\bf 46, }1671 (1992).

\bibitem{26}  S. Coleman and E. Weinberg, Phys. Rev. D {\bf 9}, 3320 (1974).

\bibitem{27}  E. W. Kolb and M. S. Turner, {\it The Early Universe (}
Addison-Wesley Publishing Company, 1990).

\bibitem{28}  R. Dominguez-Tenreiro and Mariano Quiros, {\it An introduction
to Cosmology and Particle Physics (}World Scientific, 1988), Chap. 7.

\bibitem{29}  G. Micle and P. Vitale, Nucl. Phys. B {\bf 494, }365 (1997) 
{\bf .}

\bibitem{30}  Klaus Kirsten, Class. Quantum Grav. {\bf 10, }1461 (1993).

\bibitem{31}  E. Elizalde, Yu I. Shil'nov and V. V. Chitov, Class. Quantum
Grav. {\bf 15, }735 (1998).

\bibitem{32}  T. Inagaki, T. Muta and S. D. Odintsov, Progr. Theor. Phys.
Suppl. {\bf 127, }93 (1997{\bf ). }

\bibitem{33}  A. Vilenkin and L. H. Ford, Phys. Rev. D {\bf 26}, 1231 (1982).

\bibitem{34}  D. J. O'Connor, B. L. Hu and T. C. Shen, Phys. Letts. {\bf 
130B, }31 (1983).

\bibitem{35}  M. M. Janson, Lett. Nuovo Cimento {\bf 15, }231 (1976).

\bibitem{36}  A. A. Grib and V. M. Mosteparenko, JETP Lett. {\bf 25}, 302
(1977).

\bibitem{37}  M. S. Madsen, Class. Quantum Grav. {\bf 5, }627 (1988).
\end{thebibliography}
\end{document}